\documentclass[aps,prd,twocolumn,showpacs,amsmath,amssymb,nofootinbib]{revtex4}

\newcommand{\spcpnct}{\;}                              
\newcommand{\period}{{\mbox{\spcpnct.}\relax}}         
\newcommand{\commae}{{\mbox{\spcpnct,}\relax}}         
\newcommand{\comma}{{\mbox{\spcpnct,\quad}\relax}}     

\newcommand{\tens}[1]{{#1}}
\newcommand{\grad}{{\tens{d}}}
\newcommand{\gradc}{\partial}
\newcommand{\lied}{\pounds}
\newcommand{\covd}{{\tens{\nabla}}}

\newcommand{\cv}[1]{{\tens{\partial}}_{#1}}
\newcommand{\ep}{{\tens{\epsilon}}}
\newcommand{\A}[1]{A^{\!(#1)}}
\newcommand{\Ric}{{\mathrm{Ric}}}
\newcommand{\Ricc}{{\mathrm{Ric}}}

\newcommand{\qobs}[2][]{{{}^{#1}\!K_{(#2)}}}
\newcommand{\lobs}[2][]{{{}^{#1}\!L_{(#2)}}}
\newcommand{\PKV}{{\tens{\xi}}}
\newcommand{\PKVc}{\xi}
\newcommand{\qKT}[1]{{\tens{k}_{(#1)}}}
\newcommand{\lKT}[1]{{\tens{l}_{(#1)}}}
\newcommand{\qKTc}[1]{k_{(#1)}}
\newcommand{\lKTc}[1]{l_{(#1)}}
\newcommand{\qop}[2][]{{{}^{#1}\!{\mathcal{K}_{(#2)}}}}
\newcommand{\lop}[2][]{{{}^{#1}\!{\mathcal{L}_{(#2)}}}}
\newcommand{\qsop}[2][]{{{}^{#1}\!{\tilde{\mathcal{K}}}_{(#2)}}}
\newcommand{\lsop}[1]{{{\tilde{\mathcal{L}}_{(#1)}}}}
\newcommand{\sop}[1]{{\tilde{\mathcal{X}}_{(#1)}}}
\newcommand{\qsc}[1]{\Xi_{#1}}
\newcommand{\lsc}[1]{\Psi_{#1}}
\newcommand{\qsf}[1]{{\tilde\Xi}_{#1}}
\newcommand{\lsf}[1]{{\tilde\Psi}_{#1}}

\newcommand{\EMp}{{\tens{A}}}
\newcommand{\EMpc}{A}
\newcommand{\ph}{\varphi}

\renewcommand{\tens}[1]{{\boldsymbol{#1}}}
\renewcommand{\Ric}{{\mathbf{Ric}}}

\begin{document}
\title{Charged particle in higher dimensional weakly charged rotating black hole spacetime}

\author{Valeri P. Frolov}
\email{vfrolov@ualberta.ca}

\affiliation{Theoretical Physics Institute,
University of Alberta,\\
Edmonton, Alberta, Canada T6G 2G7}

\author{Pavel Krtou\v{s}}

\email{Pavel.Krtous@utf.mff.cuni.cz}

\affiliation{Institute of Theoretical Physics,
Faculty of Mathematics and Physics, Charles University in Prague,
V~Hole\v{s}ovi\v{c}k\'ach 2, Prague, Czech Republic\\}

\date{October 11, 2010}  

\begin{abstract}
We study charged particle motion in weakly charged higher dimensional
black holes. To describe the electromagnetic field we use a test field
approximation and use the higher dimensional Kerr-NUT-(A)dS metric as
a background geometry. It is shown that for a special configuration of
the electromagnetic field the equations of motion of charged particles
are completely integrable. The vector potential of such a field is
proportional to one of the Killing vectors (called primary Killing
vector) from the `Killing tower' of symmetry generating objects which
exists in the background geometry. A free constant in the definition of
the adopted electromagnetic potential is proportional to the electric
charge of the higher dimensional black hole. The full set of independent
conserved quantities in involution is found. It is demonstrated, that
Hamilton--Jacobi equations are separable, as well as the corresponding
Klein--Gordon equation and its symmetry operators.
\end{abstract}


\maketitle

\section{Introduction}\label{sc:intro}

In this paper we describe an interesting class of spacetimes where
the equations of motion of charged particles allows a complete
separation of variables. Namely we study weakly charged rotating
higher dimensional black holes.  We assume that a background geometry
is a solution of the (vacuum)  Einstein equations and include the
electromagnetism as a test  field which does not affect the geometry.
It is well known that in  the four dimensional case this approach can
be useful. The reason is  that for known charged elementary particles
the ratio of the charge  to mass is very large. As a result, a test
electromagnetic field,  which does not change the black hole
geometry, can dramatically change  the motion of charged particles
(see e.g. \cite{AlievGaltsov:1989b,FrolovShoom:2010} and references therein).
Another application of
the test electromagnetic field approximation is study of the
gyromagnetic ratio of higher dimensional rotating black hole
\cite{AlievFrolov:2004}.

We study charged particle motion in a spacetime with a test
electromagnetic field. We focus on  the case when the background
geometry describes a  rotating higher-dimensional black hole
\cite{MyersPerry:1986}, and its generalization with `NUT' parameters
and/or with a non-trivial cosmological constant
\cite{ChenLuPope:2006,HamamotoEtal:2007}. The Kerr-NUT-(A)dS metrics
in the higher dimensions have been extensively studied recently. In
particular it was demonstrated that they have a number of
`miraculous' properties which make then similar to their
four-dimensional `cousin'. In particular, it was found that  the most
general solution of the Einstein equations with the cosmological
constant, describing higher dimensional rotating black holes with NUT
parameters (Kerr-NUT-(A)dS metric)  possesses a non-degenerate closed
2-form of the conformal Killing-Yano tensor (principal Killing-Yano
tensor) \cite{FrolovKubiznak:2007,KubiznakFrolov:2007}.
It was shown that  this object generates a
`tower' of conserved quantities
\cite{KrtousEtal:2007a,PageEtal:2007,KrtousEtal:2007b,Frolov:2008}
which makes the geodesic
equations completely integrable in these spacetimes
\cite{PageEtal:2007,KrtousEtal:2007b,HouriEtal:2008}. Later it was shown that
the Hamilton--Jacobi and Klein--Gordon  equations are separable
\cite{FrolovEtal:2007}. Analogous results on separability for another
field equations in this background have been obtained in
\cite{OotaYasui:2008a,SergyeyevKrtous:2008,OotaYasuia:2008b,KubiznakFrolov:2008,ConnellEtal:2008,KubiznakEtal:2008}.
For a general review, see \cite{KubiznakFrolov:2007}.
The presence of a principal Killing-Yano tensor imposes restrictions
on the form of the metric. Namely, the metric of the spacetime
can be written in the form, where the only freedom is a set of
functions of one variables. This result was proved in
\cite{HouriEtal:2007,HouriEtal:2008,KrtousFrolovKubiznak:2008}.

In this paper we demonstrate  that weak charges of the higher
dimensional black hole solution do  not change their remarkable
property: the equations of a charged particle motion remains
completely integrable. In the four dimensional case this  result is
not surprising: Motion of charged particles in the Kerr-Newman
spacetime has the same property \cite{Carter:1968a} and our result
can be  thus obtained by linearization. In five dimensions our
results might be related to the complete integrability of the
particle motion equations in black hole solutions of the Chern-Simon
version of Maxwell-Einstein equations \cite{ChongEtal:2005,KubiznakEtal:2009}.
In the higher dimensions the obtained results are much less trivial.

The paper is organized as follows. Section~\ref{sc:prelim}
contains some preliminary material. Required information concerning
spacetimes with a non-degenerate principal Killing-Yano tensor is
collected in  Section~\ref{sc:geometry}. Adopted ansatz for a test
electromagnetic field in this geometry is described in
Section~\ref{sc:EM}. In  Section~\ref{sc:motion},  we prove that the
motion of a charged particle in the electromagnetic field, generated by
the primary Killing vector, is complete integrable. In
Sections~\ref{sc:HJ} and \ref{sc:KG} we prove the separability of the
Hamilton--Jacobi equations and of all symmetry operators of the
Klein--Gordon operator. For simplicity we give the proofs in the even
dimensional case, but they are valid in any number of dimensions. The
changes required in the odd dimensional case are discussed in
Section~\ref{sc:odd}. The results of the paper are briefly summarized
in the last Section.

\section{Preliminaries}
\label{sc:prelim}

Consider a particle with a mass $\mu$, a charge $q$, which is moving in
the electromagnetic field ${\tens{F} = \grad\EMp}$.
Its equation of motion is
\begin{equation}\label{eqmot}
  \mu\,\frac{D^2 x^a}{d\tau^2}= q F^{a}{}_{b} \frac{Dx^b}{d\tau}\period
\end{equation}
Here, ${D/d\tau}$ is the covariant derivative with respect
to the proper time ${\tau}$.
It is useful to introduce the affine parameter $\lambda=\tau/\mu$.
Denoting by the dot a covariant derivative with respect
to parameter $\lambda$, the equation of motion can be rewritten as
\begin{equation}\label{eqmotlambda}
  \ddot{x}^a=q F^{a}{}_{b}\,\dot{x}^b\period
\end{equation}

It is well known that the symmetries
of the background geometry guarantee the existence of conserved quantities
even for motion under influence of the electromagnetic field,
provided that the electromagnetic field satisfies some consistency conditions.
Let us recall that a Killing vector~${\tens{\xi}}$ and a rank two Killing
tensor~${\tens{k}}$ satisfy the equations
\begin{equation}\label{KTcond}
\begin{gathered}
    \nabla_{\!(a}\xi_{b)}=0\commae\\
    k_{ab}=k_{(ab)} \comma \nabla_{\!(a}k_{bc)}=0\period
\end{gathered}
\end{equation}
If the spacetime possesses a Killing vector ${\tens{\xi}}$, the component
of the canonical momentum along the Killing vector, i.e., the quantity
\begin{equation}\label{linpcons}
    p_\xi = \xi^a ( g_{ab}\, \dot{x}^b + q \EMpc_a )
\end{equation}
is conserved if the vector potential ${\EMp}$ is Lie-conserved along ${\tens{\xi}}$,
\begin{equation}\label{linpconscond}
    \lied_{\tens{\xi}}\EMp = [\tens{\xi},\EMp]=0\period
\end{equation}
The component of the velocity along ${\tens{\xi}}$,
\begin{equation}\label{linucons}
    u_\xi = \xi_a \, \dot{x}^a\commae
\end{equation}
is conserved if
\begin{equation}\label{linuconscond}
    \xi^n F_{an}=0\period
\end{equation}
This second condition can be generalized to quantities
quadratic in velocities if the background geometry has
a rank two Killing tensor ${\tens{k}}$.\footnote{%
Analogy of condition \eqref{linpconscond} for quadratic
quantity generalizing \eqref{linpcons} is not so straightforward.
It involves Schouten--Nijenhuis brackets ${[\tens{k},\EMp]_{\textrm{SN}}}$,
as could be expected, but also some additional non-trivial conditions
on ${\tens{k}}$, ${\EMp}$, and their first derivatives.}
Namely, the quantity
\begin{equation}\label{quaducons}
    \dot{x}^a k_{ab}\, \dot{x}^b
\end{equation}
is conserved if\footnote{%
A related condition in terms of Killing-Yano tensor generating
${\tens{k}}$ can be found in \cite{AcikEtal:2009}.}
\begin{equation}\label{quaduconscond}
    k_{(a}{}^n \,F_{b)n} = 0\period
\end{equation}
Since the metric ${\tens{g}}$ is the Killing tensor satisfying
trivially this condition, we get obvious conservation of the norm
of the velocity ${\dot{x}^a g_{ab}\dot{x}^b}$.

It will be useful to work also with Hamiltonian formalism.
The equation of motion \eqref{eqmotlambda} follows from the Lagrangian
\begin{equation}\label{lagr}
  L=\frac12 g_{ab}\dot{x}^a\dot{x}^b + q\EMpc_{a}\dot{x}^a\period
\end{equation}
To write a Hamiltonian one defines the momentum
\begin{equation}\label{mom}
  p_{a}=\frac{\partial L}{\partial \dot{x}^a}=g_{ab}\dot{x}^b+q\EMpc_a\commae
\end{equation}
and the corresponding Hamiltonian reads
\begin{equation}\label{ham}
  H=\frac12 g^{ab}(p_a-q\EMpc_a)(p_b-q\EMpc_b)\period
\end{equation}
Since it does not depend on $\lambda$, the Hamiltonian is the integral
of motion. For our choice of the affine parameter~$\lambda$ one finds that
its value is given by
\begin{equation}\label{hamval}
  H=-\frac12\mu^2\period
\end{equation}

The conservation law \eqref{hamval} with Hamiltonian \eqref{ham} implies
the following Hamilton--Jacobi equation for the classical action
${\mathcal{S}= -\frac12 \lambda \mu^2+S(x^a)}$:
\begin{equation}\label{HJ}
    -\mu^2 = g^{ab}\bigl(\gradc_{a}S - q\EMpc_{a}\bigr)\bigl(\gradc_{b}S - q\EMpc_{b}\bigr)\period
\end{equation}
From the same Hamiltonian one obtains the equation
for a charged massive field $\ph$ by substituting
$p_a\to-i\nabla_a$. The corresponding Klein-Gordon equation is
\begin{equation}\label{KGA}
    \bigl[[\nabla_{\!a}-iq\EMpc_{a}]\,g^{ab}\,[\nabla_{\!b}-iq\EMpc_{b}]-\mu^2\bigr]\ph = 0\period
\end{equation}

Consider now a Ricci-flat spacetime, ${\Ric=0}$. In the Lorentz gauge
${\nabla_{\!n}\EMpc^n=0}$, the Maxwell equations
${\nabla_{\!n}F^{an}=0}$ reads ${\nabla_{\!n}\nabla^n\EMpc_a=0}$.
The Killing vector ${\tens{\xi}}$ obeys the same equation
${\nabla_{\!n}\nabla^n\xi_a=0}$. This means that the Killing vector
field can be used as a potential of a special test electromagnetic
field $\EMp$
\begin{equation}\label{KillEM}
   \EMp = Q\, \tens{\xi}\period
\end{equation}
Here, $Q$ is a normalization constant parameterizing the strength of the field.

Let us assume that the background spacetime is even more special,
namely, that it allows the separability of uncharged Hamilton--Jacobi
and Klein--Gordon equations. It is natural to ask, what happens
with these equations when one consider the system with the
test Killing electromagnetic field \eqref{KillEM}.

If the separation takes place with respect to
the Killing coordinate corresponding to the Killing vector ${\xi}$,
\begin{equation}\label{xisep}
  \xi^a\gradc_a S=\Psi \comma
  \xi^a\gradc_a \ph= \Psi\,\ph\commae
\end{equation}
with ${\Psi}$ being the separation constant,
the charged Hamilton--Jacobi \eqref{HJ} and Klein--Gordon equations
\eqref{KGA} take the form
\begin{gather}\label{HJKGM}
    g^{ab}\,\gradc_a S\, \gradc_b S  + M^2 = 0\commae\\
    \bigr[g^{ab}\,\nabla_{\!a}\nabla_{\!b} -M^2\bigr]\ph = 0\period
\end{gather}
Here, the function ${M^2}$ is given by
\begin{equation}\label{Mdef}
  M^2=\mu^2-2e\Psi+e^2\xi^2\commae
\end{equation}
with ${e=qQ}$.

These equations clearly resemble the uncharged case.
Thus in the presence of the test Killing electromagnetic field the Hamilton--Jacobi
and Klein--Gordon equations preserve their form with the only change
of the constant ${\mu^2}$ by a function ${M^2}$. Evidently, the constant
shift ${-2e\Psi}$ does not affect the complete separability property
of the initial equations. Non-trivial obstacle to the separability can
create the term ${\xi^2}$. We shall describe now a physically
interesting case when the complete separability is not broken by the
external Killing electromagnetic field.

\section{Higher-dimensional black hole geometry}
\label{sc:geometry}

Rotating black hole solution in higher dimension belong
to broader class of spacetimes studied in
\cite{ChenLuPope:2006,HamamotoEtal:2007,HouriEtal:2007,KrtousFrolovKubiznak:2008}.
In even dimension ${D=2n}$, geometry of such spacetimes is described by the metric
\begin{equation}\label{metric}
\tens{g}
  =\sum_{\mu=1}^n\;\biggl[\; \frac{U_\mu}{X_\mu}\,{\grad x_{\mu}^{\;\,2}}
  +\, \frac{X_\mu}{U_\mu}\,\Bigl(\,\sum_{j=0}^{n-1} \A{j}_{\mu}\grad\psi_j \Bigr)^{\!2}
  \;\biggr]
  \period
\end{equation}
Here ${x_\mu}$, ${\mu=1,\dots,n}$, correspond to radial and `azimuthal'
directions and ${\psi_k}$, ${k=0,\dots,n-1}$ to temporal and longitudinal
directions, namely ${\psi_0=t}$. The radial coordinate and some
other quantities are rescaled by the imaginary unit ${i}$ in order
to bring the metric into a more symmetric form, cf.\ e.g.~\cite{ChenLuPope:2006};
however, the metric is real. The signature of the metric depends on
the signs of the metric functions. We use Latin indices from the
beginning of the alphabet to label the
whole coordinate set: ${\{x^a\}=\{x_\mu,\psi_k\}}$.

The functions ${U_\mu}$ and ${\A{k}_\mu}$ are defined as follows
\begin{equation}\label{UAdef}
  \A{k}_{\mu}=\hspace{-5mm}
    \sum\limits_{\substack{\nu_1,\dots,\nu_k=1\\\nu_1<\dots<\nu_k,\;\nu_i\ne\mu}}^n\!\!\!\!\!
    x^2_{\nu_1}\cdots\ x^2_{\nu_k}\comma
  U_{\mu}=\prod\limits_{\substack{\nu=1\\\nu\ne\mu}}^{n}(x_{\nu}^2-x_{\mu}^2)\period
\end{equation}
These functions satisfy the important relations \cite{FrolovEtal:2007}
\begin{equation}\label{AUrel}
\hspace*{-1mm}  \sum_{\mu=1}^n\! \A{i}_\mu
\frac{(-x_\mu^2)^{n\!-\!1\!-\!j}}{U_\mu} =
  \delta^i_j\comma\!\!
  \sum_{j=0}^{n-1}\!\A{j}_\mu \frac{(-x_\nu^2)^{n\!-\!1\!-\!j}}{U_\nu} = \delta^\nu_\mu\commae
\end{equation}
for ${i,j=0,\dots,n-1}$ and ${\mu,\nu=1,\dots,n}$.

The quantities ${X_\mu}$, ${\mu=1,\dots,n}$, are functions of a single variable,
that is each ${X_\mu}$ depends only on the variable ${x_\mu}$, ${X_\mu=X_\mu(x_\mu)}$.
However, when these functions are not specified,
the metric \eqref{metric} does not satisfy the
vacuum Einstein equations. Without choosing a concrete form
of functions ${X_\mu}$ we speak about, so called, ``off-shell'' geometry.
The vacuum (with a cosmological constant) black hole geometry is recovered
\cite{ChenLuPope:2006,HamamotoEtal:2007} by setting
\begin{equation}\label{BHXs}
  X_\mu = b_\mu\, x_\mu + \sum_{k=0}^{n}\, c_{k}\, x_\mu^{2k}\period
\end{equation}
The constants ${c_k}$ and ${b_\mu}$ are then related to angular momenta,
mass, NUT parameters, and the cosmological constant (which is proportional
to~${c_n}$).

We can write the metric \eqref{metric} in the diagonal form
\begin{equation}\label{diagmetric}
\tens{g}= \sum_{\mu=1}^n\,
    \biggl(\,\frac{U_\mu}{X_\mu}\,\ep^\mu  \ep^\mu
    + \frac{X_\mu}{U_\mu}\,\ep^{\hat \mu}  \ep^{\hat\mu}\,\biggr)
  \period
\end{equation}
introducing the non-normalized one-forms
${\{\ep^\mu,\ep^{\hat\mu}\}}$,
\begin{equation}\label{formframe}
  \ep^\mu = \grad x_{\mu}\comma
  \ep^{\hat\mu} = \sum_{j=0}^{n-1}\A{j}_{\mu}\grad\psi_j
  \period
\end{equation}

In this frame, the Ricci tensor for the off-shell geometry is diagonal,
\begin{equation}\label{Ric}
    \Ric = \sum_{\mu=1}^n r_\mu
    \biggl(\,\frac{U_\mu}{X_\mu}\,\ep^\mu  \ep^\mu
    + \frac{X_\mu}{U_\mu}\,\ep^{\hat \mu}  \ep^{\hat\mu}\,\biggr)\commae
\end{equation}
where
\begin{equation}\label{rmu}
    r_\mu = - \frac1{2x_\mu}\biggl[\sum_{\nu=1}^n
    \frac{x_\nu^2\bigl(x_\nu^{-1} X_\nu\bigr){}_{,\nu}}{U_\nu}\biggr]_{\!,\mu}\period
\end{equation}
For the Einstein spacetime, polynomials \eqref{BHXs} lead to a constant value
${r_\mu = -(2n-1)c_n= \Lambda/{(n-1)}}$.

The off-shell geometry \eqref{metric} is endowed with a lot of symmetries. The symmetry set,
forming a `Killing tower', is generated by a single object called a
\emph{principal conformal Killing-Yano tensor}
\cite{KubiznakFrolov:2007,KrtousEtal:2007a}. This is a non-degenerate
closed conformal Killing-Yano 2-form $\tens{h}$,
\begin{equation}\label{PKY}
    \tens{h} = \sum_{k=1}^n x^\mu\, \ep^\mu\wedge\ep^{\hat\mu}\period
\end{equation}
The explicit symmetries are encoded by the Killing vectors ${\lKT{k}}$,
${k=0,\dots,n-1}$,
\begin{equation}\label{lKT}
\lKT{k}=\cv{\psi_k}\period
\end{equation}
The geometry possesses also hidden symmetries encoded by the 2nd rank
Killing tensors ${\qKT{j}}$, ${j=0,\dots,n-1}$ which in their covariant
form read
\begin{equation}\label{qKT}
\qKT{j}=\sum_{\mu=1}^{n}\A{j}_\mu
\biggl(\frac{U_\mu}{X_\mu}\,\ep^\mu   \ep^\mu +
\frac{X_\mu}{U_\mu}\,\ep^{\hat \mu}\ep^{\hat \mu}\biggr)\period
\end{equation}
In particular, for ${j=0}$, the Killing tensor reduces
to the metric, ${\qKTc{0}{}_{ab}=g_{ab}}$.

The Killing vector ${\PKV}$,
\begin{equation}\label{PKVfromPKY}
  \PKVc^a=\frac1{D-1}\nabla_{\!n}h^{na}\commae
\end{equation}
is called \emph{primary Killing vector}, since it turns out
that it is the first in the tower of the Killing vectors defined above,
${\PKV=\lKT{0}=\cv{\psi_0}}$.
All Killing vectors can be actually obtained from the primary Killing vector
using, for our purposes, important relations
\begin{equation}\label{lKTqKTrel}
   \lKTc{k}^a=\qKTc{k}^{an}\, \PKVc_{n}\period
\end{equation}

\section{Test electromagnetic field}
\label{sc:EM}

In higher dimensions, a generalization of the above described
geometry to the case of arbitrary rotating charged black hole
is not known. However, we can investigate at least the weakly
charged black hole, i.e., the neutral black hole spacetime
with a test electromagnetic filed satisfying the Maxwell equations
in such a background. Such an approximation is plausible since
even the electromagnetic field small enough not to influence
the background geometry can cause significant changes in the
particle motion thanks to a large charge-to-mass ration ${q/\mu}$
for typical particles.

Test electromagnetic fields on the background \eqref{metric}
have been studied, e.g., in \cite{Krtous:2007}. However, in this
paper we concentrate on the special Killing electromagnetic field,
i.e., we assume the electromagnetic field with the vector
potential \eqref{KillEM} given by a Killing vector ${\tens{\xi}}$.
In general, the electric current generating such a field is
${ J^a = 2 Q\, \Ricc^{a}{}_{b} \xi^b }$
For Ricci-flat spacetimes the Killing electromagnetic field is thus
source-free, and for the Einstein spaces the electric current
is aligned along the generating Killing vector.

In the following we will investigate the electromagnetic field
generated by the primary Killing vector ${\PKV}$ on the black
hole background \eqref{metric}. Since we are
interested in the motion of the particle with charge ${q}$, we
use the following parametrization:
\begin{equation}\label{EMpot}
    q\EMp = e\,\PKV\period
\end{equation}
The constant ${e/q}$ parametrize the field strength and it is
proportional to the test charge of the black hole.

The results derived in the following sections do not depend
on a nature of the source ${\tens{J}}$ of the primary Killing
electromagnetic field. They hold for a general off-shell
geometry \eqref{metric}. In such a general case, the
corresponding electric current is
\begin{equation}\label{PKVJ}
    q \tens{J}= 2e\sum_{\mu=1}^n r_\mu \ep_{\hat\mu}\period
\end{equation}
It represents the source distributed, in general, in the whole spacetime,
which is not very reasonable. Therefore, physically the most interesting case
is when the vacuum Einstein equations ${\Ric=0}$ are satisfied, so
that ${\tens{J}=0}$.

In this case, the electromagnetic field \eqref{EMpot}
belongs to the class of the electromagnetic fields studied in
\cite{Krtous:2007}. Namely, the potential \eqref{EMpot}
is gauge equivalent to the choice ${e_\mu=e/q\, b_\mu}$
of the constants ${e_\mu}$ parameterizing the field in \cite{Krtous:2007},
with ${b_\mu}$ from \eqref{BHXs}. In four dimension, it can be also
obtained from the electromagnetic field of the Kerr-Newman solution
by linearization.

\section{Phase space description of the particle motion}
\label{sc:motion}

The motion of the particle in the spacetime $\mathcal{M}$
with the metric \eqref{metric} can be described in phase space represented as
the cotangent bundle $\mathbf{T}^*\mathcal{M}$.
The basic variable is the one-form of canonical momenta ${\tens{p}}$,
components $p_a$ of which are canonically conjugate to
$x^a$, $a=1,\dots,D$.

The motion in the absence of the electromagnetic field was studied
in \cite{PageEtal:2007,KrtousEtal:2007b} and it was shown that
the Killing vectors \eqref{lKT} and tensors \eqref{qKT} generate functionally
independent and mutually Poisson-commuting observables which are
linear and quadratic in momentum ${\tens{p}}$, namely,
\begin{equation}\label{lqobs0}
  \lobs[0]{k}=\lKTc{k}^a\, p_a\comma
  \qobs[0]{j}=\qKTc{j}^{ab}\, p_a p_b\period
\end{equation}
The commutations relations of the observables \eqref{lqobs0} are equivalent
to nontrivial geometrical relations among the Killing vectors and tensors:\footnote{%
These relation correspond to vanishing Schouten--Nijenhuis brackets
among all tensors ${\lKT{k}}$ and ${\qKT{j}}$.},
\begin{gather}
    \lKTc{k}^n\nabla_{\!n}\lKTc{l}^a - \lKTc{l}^n\nabla_{\!n}\lKTc{k}^a=0 \commae\label{LLident}\\
    \lKTc{k}^n\nabla_{\!n}\qKTc{j}^{ab} = \qKTc{j}^{an}\,\nabla_{\!n}\lKTc{k}^{b}+\qKTc{j}^{bn}\,\nabla_{\!n}\lKTc{k}^{a}\commae\label{LKident}\\
    \qKTc{i}^{n(a}\,\nabla_{\!n}\qKTc{j}^{bc)}-\qKTc{j}^{n(a}\,\nabla_{\!n}\qKTc{i}^{bc)}=0\period\label{KKident}
\end{gather}

The geodesic motion of an uncharged particle is generated by
the Hamiltonian ${{}^0\!H}$ which is essentially one of these observables,
\begin{equation}\label{Hamilt0}
    {}^0\!H=\frac12\,\qobs[0]{0}=\frac12\,p_a g^{ab} p_b\period
\end{equation}
All the observables ${\lobs[0]{k}}$ and ${\qobs[0]{j}}$ are thus
conserved quantities (i.e., integrals of motion) for  the geodesic
motion. They are  independent and in involution.  As a
result, according to the Liouville theorem  the geodesic motion is
completely integrable.

However, in this paper we want to investigate a motion of
a charged particle modified  by
the electromagnetic field \eqref{EMpot}.
Such a motion is described by the Hamiltonian \eqref{ham}, i.e.,
\begin{equation}\label{Hamilte}
    {}^{e}\!H=\frac12 (p_a-e\PKVc_a)\, g^{ab}\, (p_b-e\PKVc_b)\period
\end{equation}
We demonstrate now that the corresponding equations also complete integrable.
However, in this case the integrals of motion must be modified. Namely,
we define new observables
\begin{equation}\label{lqobs}
  \lobs[e]{k}=\lKTc{k}^a\, p_a\comma
  \qobs[e]{j}=\qKTc{j}^{ab}\, (p_a-e l_a) (p_b-e l_b)\period
\end{equation}
We show that these observables are  in involution
\begin{equation}\label{commobs}
    \{ \lobs[e]{l}, \lobs[e]{k}\} {=} 0\;,\;\;
    \{ \lobs[e]{l}, \qobs[e]{j}\} {=} 0\;,\;\;
    \{ \qobs[e]{i}, \qobs[e]{j}\} {=} 0\period
\end{equation}
And since ${{}^e\!H=\frac12\qobs[e]{0}}$, they form a complete
set of conserved quantities for motion of the particle under an
influence of the electromagnetic field.

The modified observables \eqref{lqobs} can be related to the
observables \eqref{lqobs0} as
\begin{equation}\label{obsrel}
\begin{gathered}
    \lobs[e]{k} = \lobs[0]{k} \equiv \lobs{k}\commae\\
    \qobs[e]{j} = \qobs[0]{j} -2 e\, \lobs{j} + e^2 \qKTc{j}^{ab}\,\PKVc_{a}\PKVc_{b}\period
\end{gathered}
\end{equation}
After plugging these expressions into equations \eqref{commobs} and using the fact
that quantities \eqref{lqobs0} commute with each other, it remains to prove that
${\{\lobs{k},\qKTc{j}^{ab}\,\PKVc_{a}\PKVc_{b}\}=0}$
and
${\{\qobs[0]{i},\qKTc{j}^{ab}\,\PKVc_{a}\PKVc_{b}\}+\{\qKTc{i}^{ab}\,\PKVc_{a}\PKVc_{b},\qobs[0]{j}\}=0}$.
Evaluating the Poisson brackets\footnote{%
In covariant formalism we have
${\{A,B\}=\nabla_{\!n}A\;\partial^n\! B-\partial^n \!A\; \nabla_{\!n} B}$,
where ${\nabla_{\!n}}$ is the covariant derivative ``ignoring'' the momentum
dependence of the phase space observables ${A}$, ${B}$ (taking a parallelly
transported ${\tens{p}}$ as a constant) and ${\partial^n}$ is derivative
with respect to the momentum ${\tens{p}}$. Cf., for example,
the Appendix of~\cite{KrtousEtal:2007b}.}
we can translate these equalities to the language of tensors on the spacetime
\begin{equation}\label{toprove}
\begin{gathered}
   \lKTc{k}^n\nabla_{\!n}\bigl(\qKTc{j}^{ab}\, \PKVc_a\PKVc_b\bigr) =0\commae\\
   \qKTc{i}^{cn}\nabla_{\!n}\bigl(\qKTc{j}^{ab}\, \PKVc_a\PKVc_b\bigr)
   -\qKTc{j}^{cn}\nabla_{\!n}\bigl(\qKTc{i}^{ab}\, \PKVc_a\PKVc_b\bigr)=0\period
\end{gathered}
\end{equation}

The first equality can be easily proved realizing that thanks to \eqref{lKTqKTrel}
${\qKTc{j}^{ab}\, \PKVc_a\PKVc_b= g_{ab}\, \lKTc{j}^a \PKVc^{b}}$ and
all the quantities ${\lKT{j}}$, ${\PKV}$, and ${\tens{g}}$
are Lie-conserved along ${\lKT{k}}$.

In the second equality one has to perform the covariant derivatives
on the tensor products in their argument and use the identity
\begin{equation}\label{knablakidentity}
\begin{split}
    &\qKTc{i}^{cn}\nabla_{\!n}\qKTc{j}^{ab}-\qKTc{j}^{cn}\nabla_{\!n}\qKTc{i}^{ab} =\\
         &{=}{-}\Bigl(\qKTc{i}^{an}\nabla_{\!n}\qKTc{j}^{bc}{+}\qKTc{i}^{bn}\nabla_{\!n}\qKTc{j}^{ac}\Bigr)
         {+}\Bigl(\qKTc{j}^{an}\nabla_{\!n}\qKTc{i}^{bc}{+}\qKTc{j}^{bn}\nabla_{\!n}\qKTc{i}^{ac}\Bigr)\commae
\end{split}\raisetag{35pt}
\end{equation}
which follows from Eq.~\eqref{KKident}. Substituting
${\bigl(\nabla_{\!n} \qKTc{j}^{ac}\bigr)\PKVc_a}={\nabla_{\!n}\lKTc{j}^c-\bigl(\nabla_{\!n}\PKVc_a\bigr)\qKTc{j}^{ac}}$
 (cf.~\eqref{lKTqKTrel}), we finally get
\begin{equation}\label{proof}
\begin{split}
    &2\qKTc{i}^{ca}\bigl(\nabla_{\!a}\PKVc_{b}\bigr)\qKTc{j}^{bn}\PKVc_n
    +     2\qKTc{j}^{ca}\bigl(\nabla_{\!a}\PKVc_{b}\bigr)\qKTc{i}^{bn}\PKVc_n\\
    &\qquad- (\text{terms with $i$ and ${j}$ exchanged}) = 0\period
\end{split}
\end{equation}
At the end we used the definition \eqref{KTcond} of the Killing vector.

We thus concluded the proof that the observables \eqref{lqobs}
are in involution and therefore they also commute
with Hamiltonian. They are functionally independent, which follows
from the independence of variables \eqref{lqobs0} which was proven
in \cite{PageEtal:2007,KrtousEtal:2007a}. Therefore, the Hamiltonian
describe completely integrable motion with linear and quadratic
integrals of motion.

\section{Hamilton-Jacobi equations}
\label{sc:HJ}

Alternatively, instead of the phase space descriptions we can use
Hamilton--Jacobi theory to describe the particle motion.
Namely, for each conserved quantity we can write down the
Hamilton--Jacobi equation for the Hamilton--Jacobi function
(classical action) ${S}$. It is obtained by substituting ${\grad S}$
for the momentum ${\tens{p}}$ in definitions of conserved quantities:
\begin{gather}
    \lKTc{k}^a\, \gradc_{a} S = \lsc{k}\commae\label{lHJeqs}\\
    \qKTc{j}^{ab}\; (\gradc_{a}S-e\,\PKVc_{a}) (\gradc_{b} S-e\,\PKVc_{b}) = \qsc{j}\period\label{qHJeqs}
\end{gather}
Here, ${\lsc{k}}$ and ${\qsc{j}}$ are constants of the motion.

Now we show that all these equations can be simultaneously solved
by the separability ansatz for ${S}$
\begin{equation}\label{sepS}
    S = \sum_{\mu=1}^n S_\mu(x_\mu) + \sum_{k=0}^{n-1} \lsc{k}\psi_k\commae
\end{equation}
where the functions ${S_\mu(x_\mu)}$ are functions of a single
variable ${x_\mu}$.

For ${e=0}$ such a separability of the Hamilton--Jacobi equations was
proved in \cite{FrolovEtal:2007} and \cite{SergyeyevKrtous:2008}.
Adding electromagnetic field, generated by the primary Killing
vector, does not change the first set of equations \eqref{lHJeqs}. It
is trivially solved by the ansatz \eqref{sepS}. The equations
quadratic in ${\grad S}$ can be written as
(cf.\ Eq.~\eqref{obsrel})
\begin{equation}\label{qHJexp}
   \qKTc{j}^{ab}\; \gradc_{a}S\, \gradc_{b} S
     - 2e\,\PKVc^{a}\gradc_a S + e^2 \qKTc{j}^{ab}\PKVc{_a}\PKVc{_b} =
     \qsc{j}\period
\end{equation}
Plugging in the separability ansatz \eqref{sepS} and using the relation
\begin{equation}\label{lkl}
    \qKTc{j}^{ab}\,\PKVc_a\PKVc_b = \sum_{\mu=1}^n\frac{\A{j}_\mu}{U_\mu}X_\mu\commae
\end{equation}
one obtains
\begin{equation}\label{HJeqcoor}
\begin{split}
    \sum_{\mu=1}^n \frac{\A{j}_\mu}{U_\mu}X_\mu
       &\biggl(S_\mu'^2 + X_\mu^{-2}\Bigl(
       \sum_{k=0}^{n-1}\lsc{k}\bigl(-x_\mu^2\bigr)^{n{-}1{-}k}
       \Bigr)^2+e^2\biggr) =\\
       &= \qsc{j}+2e\lsc{j}\commae
\end{split}\raisetag{15pt}
\end{equation}
where, the prime denotes the derivative of ${S_\mu}$ with
respect to its single argument. We multiply both sides
of Eq.~\eqref{HJeqcoor} by ${(-x_\mu^2)^{n-1-j}}$ and
sum over ${j}$. Since Eq.~\eqref{AUrel} tell us
that the matrix ${(-x_\mu^2)^{n{-}1{-}j}}$
is inverse to ${\A{j}_\mu/U_\mu}$, we obtain
\begin{equation}\label{sepeqs}
    X_\mu\Bigl(S_\mu'^2 + X_\mu^{-2}\lsf{\mu}^2+e^2\Bigr) =
    \qsf{\mu}+2e\,\lsf{\mu}\period
\end{equation}
Here we introduced the polynomial functions ${\lsf{\mu}}$ and
${\qsf{\mu}}$ of one variable ${x_\mu}$ with coefficients given
by ${\lsc{k}}$ and ${\qsc{j}}$, respectively:
\begin{equation}\label{lqsfc}
    \lsf{\mu} = \sum_{k=0}^{n-1} \lsc{k}(-x_\mu^2)^{n-1-k} \comma
    \qsf{\mu} = \sum_{k=0}^{n-1} \qsc{k} (-x_\mu^2)^{n-1-k} \period
\end{equation}
Eq.~\eqref{sepeqs} gives an ordinary differential
equation for each ${S_\mu}$
\begin{equation}\label{Seq}
    \bigl(S_\mu'\bigr)^2 = \frac{\qsf{\mu}}{X_\mu}-\Bigl(\frac{\lsf{\mu}}{X_\mu}-e\Bigr)^2\period
\end{equation}
Hence, the functions ${S_\mu}$, satisfying these equations, generate
through \eqref{sepS} the solution of all Hamilton--Jacobi equations.

\section{Separability and symmetry operators of Klein--Gordon equation}
\label{sc:KG}

A field analogue of the spinless particle motion is a scalar field
governed by the Klein--Gordon equation. In the presence of the
electromagnetic field it must be modified into \eqref{KGA}.
For the electromagnetic field \eqref{EMpot} we thus get
\begin{equation}\label{KGEM}
    \bigl[[\nabla_{\!a}-ie\PKVc_{a}]\,g^{ab}\,[\nabla_{\!b}-ie\PKVc_{b}]-\mu^2\bigr]\ph = 0\period
\end{equation}
The Klein-Gordon operator has been obtained from the Hamiltonian by
the substitution ${\tens{p}\to-i\covd}$. In a similar manner we introduce
a set of operators generated by observables \eqref{lqobs}
\begin{equation}\label{lqops}
\begin{gathered}
    \lop[e]{k} = -i \lKTc{k}^a\,\nabla_{\!a}\comma\\
    \qop[e]{j} = - [\nabla_{\!a}-ie\,\PKVc_{a}]\,\qKTc{j}^{ab}\,[\nabla_{\!b}-ie\,\PKVc_{b}]\commae
\end{gathered}
\end{equation}
which turns out to be commuting with each other.
\begin{equation}\label{symopcomm}
    [\lop[e]{k},\lop[e]{k}]=0\comma
    [\lop[e]{k},\qop[e]{j}]=0\comma
    [\qop[e]{i},\qop[e]{j}]=0\period
\end{equation}
Since the charged Klein--Gordon operator in \eqref{KGEM} is simply
related to ${\qop[e]{0}}$, it also means that these operators are
symmetry operators of this Klein--Gordon operator.

In the absence of the electromagnetic field the commutation
relation \eqref{symopcomm}  were
proved in \cite{SergyeyevKrtous:2008}. The electromagnetic field
modifies only operators ${\qop[e]{j}}$ and we can write
\begin{equation}\label{opsexp}
\begin{gathered}
    \lop[e]{k} = \lop[0]{k} \equiv \lop{k}\commae\\
    \qop[e]{j} = \qop[0]{j} - 2e\,\lop{j} +
    e^2\, \qKTc{j}^{ab}\,\PKVc_{a}\PKVc_{b}\period
\end{gathered}
\end{equation}
Here we used \eqref{lKTqKTrel} and that the Killing vectors have
vanishing divergence, ${\nabla_{\!a}\lKTc{j}^a=0}$.

We can write operators \eqref{lqops} as a linear combination of
another of operators
${\lsop{k}}$ and ${\qsop[e]{j}}$:
\begin{equation}\label{lqopdual}
    \lop{k}=\sum_{\mu=1}^n\frac{\A{k}_\mu}{U_\mu}\lsop{\mu}\comma
    \qop[e]{j}=\sum_{\mu=1}^n\frac{\A{j}_\mu}{U_\mu}\qsop[e]{\mu}\commae
\end{equation}
with
\begin{equation}\label{sepop}
\begin{gathered}
  \lsop{\mu} = \sum_{j=0}^{n-1}(-x_\mu^2)^{n-1-j}\,\lop{j} \commae\\
  \qsop[e]{\mu} = \sum_{j=0}^{n-1}(-x_\mu^2)^{n-1-j}\,\qop[e]{j}\period
\end{gathered}
\end{equation}
It was shown in \cite{SergyeyevKrtous:2008}, that the
operators  ${\qsop[0]{j}}$ have a form:
\begin{equation}\label{qsopexpl0}
  \qsop[0]{\mu}
     =\Bigl[\sop{j}+ \frac{1}{X_\mu}\lsop{j}^2\Bigr]\commae
\end{equation}
with
\begin{equation}\label{xop}
  \sop{\mu}= - \frac{\partial}{\partial x_\mu}\!\biggl[X_\mu\frac{\partial}{\partial x_\mu}\biggr]\period
\end{equation}
The relations \eqref{opsexp} and \eqref{lkl} allow us to rewrite
the modified operators ${\qsop[e]{j}}$ in a similar way
\begin{equation}\label{qsopexpl}
    \qsop[e]{\mu}=\Bigl[\sop{\mu}+ \frac{1}{X_\mu}\bigl[\lsop{\mu}
    -eX_\mu\bigr]^2\Bigr] \period
\end{equation}

None of the above operators depend on the Killing coordinates
${\psi_k}$. Operators with a label $\mu$, ${\qsop[e]{\mu}}$,
${\lsop{\mu}}$, and ${\sop{\mu}}$, besides ${\partial/{\partial\psi_k}}$ depend
only on the corresponding coordinate ${x_\mu}$ and the derivative
$\partial/\partial{x_{\mu}}$. They do not contain ${x_\nu}$ or
${\partial/\partial{x_\nu}}$ for ${\nu\neq\mu}$.
As a result the operators commute among themselves
\begin{equation}\label{sopcomm}
    [\lsop{\mu},\lsop{\nu}]=0\comma
    [\lsop{\mu},\qsop[e]{\nu}]=0\comma
    [\qsop[e]{\mu},\qsop[e]{\nu}]=0\commae
\end{equation}
for ${\mu\neq\nu}$. Using \eqref{sepop} and the fact, that the
coefficients in front of the operators depends just on ${x_\mu}$, a
simple argument shows that \eqref{sopcomm} implies the commutation
\eqref{symopcomm}, cf.~\cite{SergyeyevKrtous:2008}.

Having the set of commuting operators, we can look for common
eigenfunctions.
\begin{equation}\label{eigenprob}
   \lop{k}\ph = \lsc{k}\ph\comma
   \qop[e]{j}\ph = \qsc{j}\ph
\end{equation}
with eigenvalues ${\lsc{k}}$ and ${\qsc{j}}$.
These eigenfunctions can be found using
the separability ansatz \cite{FrolovEtal:2007}
\begin{equation}\label{opsepans}
    \ph = \prod_{\mu=1}^n R_\mu(x_\mu)
    \prod_{k=0}^{n-1}\exp\bigl(i\lsc{k}\psi_k\bigr)\commae
\end{equation}
where the functions ${R_\mu(x_\mu)}$ depend on a single
variable~${x_\mu}$. Indeed, substituting \eqref{opsepans}
into \eqref{eigenprob}, the equations for ${\lop{k}}$ are
trivially satisfied and the equations for ${\qop[e]{j}}$ give
\begin{equation}\label{qeigenprob}
    \sum_{\nu=1}^n \frac{\A{j}_\nu}{U_\nu}
      \biggl(\frac{1}{R_\nu}\sop{\nu}R_\nu +
      \frac{1}{X_\nu}\bigl(\lsf{\nu}-eX_\nu\bigr)^2\biggr) = \qsc{j}
\end{equation}
Here ${\lsf{\mu}}$ (and ${\qsf{\mu}}$ below) are again given by \eqref{lqsfc}.
Summing these equations with coefficients ${(-x^2_\mu)^{n-1-j}}$ leads to
equivalent set of conditions:
\begin{equation}\label{sepcond}
    \bigl(X_\mu R_\mu'\bigr)' + \Bigl(\qsf{\mu} -
    \frac{1}{X_\mu}\bigl(\lsf{\nu}-eX_\nu\bigr)^2\Bigr)R_\mu =0\period
\end{equation}
These are ordinary differential equations in variable ${x_\mu}$
for the functions ${R_\mu}$ which guarantee that \eqref{opsepans}
solves the eigenvalue problem \eqref{eigenprob}.

In particular, for ${\qop[e]{0}}$ we obtain the separability of the Klein--Gordon equation,
which was, in the absence of the electromagnetic filed, shown in \cite{FrolovEtal:2007}.

It is straightforward to check that the semiclassical (geometrical-optic)
approximation of the eigenvalue conditions \eqref{eigenprob} leads
to the Hamilton--Jacobi equations \eqref{lHJeqs} and \eqref{qHJeqs}, where
we have to identify
\begin{equation}\label{sclaprx}
    \ph = \exp(iS)\comma\text{i.e.,}\quad R_\mu=\exp(iS_\mu)\period
\end{equation}

\section{Odd spacetime dimensions}
\label{sc:odd}

Till now we considered even dimensional spacetimes. However the
obtained results are valid in any number of dimensions. This can be
easy shown by performing similar calculations. Only some of the
equations has to be slightly changed. Here we present a short
overview of the required changes.

In spacetime dimension ${D=2n+1}$, there exists an additional angular
coordinate ${\psi_n}$ and the metric \eqref{metric} contains additional term
\cite{ChenLuPope:2006,HamamotoEtal:2007,HouriEtal:2007,KrtousFrolovKubiznak:2008}.
\begin{equation}\label{oddmetric}
\begin{split}
\tens{g}
  =&\sum_{\mu=1}^n\;\biggl[\; \frac{U_\mu}{X_\mu}\,{\grad x_{\mu}^{\;\,2}}
  +\, \frac{X_\mu}{U_\mu}\,\Bigl(\,\sum_{j=0}^{n-1} \A{j}_{\mu}\grad\psi_j \Bigr)^{\!2}
  \;\biggr]\\
  &\qquad+\frac{c}{\A{n}}\Bigl(\sum_{k=0}^n \A{k}\grad\psi_k\!\Bigr)^{\!2}
  \period
\end{split}
\end{equation}
Here, ${\A{k}}$ is defined as
\begin{equation}
  \A{k}=\!\!\!\!\!\sum\limits_{\substack{\nu_1,\dots,\nu_k=1\\\nu_1<\dots<\nu_k}}^n\!\!\!\!\!
    x^2_{\nu_1}\cdots\ x^2_{\nu_k}\commae
  \label{Adef}
\end{equation}
and ${c}$ is an auxiliary constant which can be modified by a coordinate
transformation.

The geometry has an additional Killing vector ${\lKT{n}}$
(given again by \eqref{lKT}), but the same 2nd rank Killing tensors.
The additional symmetry generates additional conserved quantity ${\lobs{n}}$
linear in momentum. It clearly Poisson-commutes with all other conserved observables.
Similarly we have additional symmetry operator ${\lop{n}}$ commuting with
other symmetry operators.

The separability ansatz for the Hamilton-Jacobi equations \eqref{sepS}
and for the eigenvalue problem of the symmetry operators \eqref{opsepans}
changes just by including term ${\lsc{n}\psi_n}$. The ordinary differential
equations for ${S_\mu}$ and ${R_\mu}$, however, acquire non-trivial additional terms
which can be partially hidden into redefinition of the polynomials ${\lsf{\mu}}$
and ${\qsf{\mu}}$. Namely, ${S_\mu}$ and ${R_\mu}$ have to satisfy
\begin{gather}\label{SRoddcond}
    \bigl(S_\mu'\bigr)^2 = \frac{\qsf{\mu}}{X_\mu}
       -\Bigl(\frac{\lsf{\mu}}{X_\mu}-e\Bigr)^2\commae\\
    \bigl(X_\mu R_\mu'\bigr)' + \frac{X_\mu}{x_\mu} R_\mu'
       + \Bigl(\qsf{\mu} - \frac{1}{X_\mu}\bigl(\lsf{\nu}-
       eX_\nu\bigr)^2\Bigr)R_\mu =0\commae\notag
\end{gather}
with
\begin{equation}\label{lqsfcodd}
    \lsf{\mu} = \sum_{k=0}^n \lsc{k}(-x_\mu^2)^{n-1-k} \comma
    \qsf{\mu} = \sum_{k=0}^n \qsc{k} (-x_\mu^2)^{n-1-k} \commae
\end{equation}
where we set ${\qsc{n} = c^{-1}\lsc{n}^2}$
(cf.~\cite{FrolovEtal:2007,SergyeyevKrtous:2008}).

\section{Summary}
\label{sc:summary}

To summarize, we proved that the dynamical equations for a charged
particle in a weakly charged Kerr-NUT-(A)dS spacetime are completely
integrable. We also demonstrated the Hamilton--Jacobi and
Klein--Gordon equations are completely separable in such a space. The
proof essentially used the remarkable properties of the geometry,
namely the existence of the principal conformal Killing-Yano tensor,
which generates the `Killing tower' of symmetries. It should be
emphasized, that the developed formalism works only for the test
electromagnetic field generated by the primary Killing vector. The
test fields connected with other Killing vectors do no possesses these
nice properties.

Let us make some general remarks, connected with our results.
Complete integrability of dynamical equations is quite a rare case.
Liouville integrability implies that a solution can be written by
applying a finite number of steps which include algebraic operations
and integration. In such a case the  phase space is regularly
foliated by trajectories. The geodesic motion in the Kerr-NUT-(A)dS
spacetimes is a new physically interesting example of completely
integrable systems. In this paper we demonstrated that these nice
properties remain valid if one includes a special type of test
electromagnetic field generated by a primary Killing vector. This
generalization allows one to study  motion of charged particles in
the weakly charged higher dimensional black holes. The results might
also be interesting for `physical' applications, for example, for
study the Hawking radiation of charged rotating black holes in higher
dimensions. They might also give some hints for the search of more
general, possibly self-consistent solutions of electrovacuum Einstein
equations, and their supersymmetric generalizations.


\section*{Acknowledgments}
V.F. thanks the Natural Sciences and Engineering
Research Council of Canada and the Killam Trust for the financial
support. P.K. was kindly supported by by Grant No. GA\v{C}R-202/08/0187,
Project No.~MSM0021620860 and appreciates the hospitality of the Theoretical
Physics Institute of the University of Alberta.
The authors would like to thank David Kubiz\v{n}\'{a}k for valuable
comments.




\end{document}